# Free-Electron Ramsey-Type Interferometry for Enhanced Amplitude and Phase Imaging of Nearfields

Tomer Bucher[1,2], Ron Ruimy[1,2], Shai Tsesses[1,3], Raphael Dahan[2], Guy Bartal[1], Giovanni Maria Vanacore[4] and Ido Kaminer[1,2]

[1]Andrew and Erna Viterbi department of Electrical & Computer Engineering, Technion-Israel Institute of Technology, 3200003, Haifa, Israel
[2]Solid State Institute, Technion-Israel Institute of Technology, Haifa 3200003, Israel
[3]Department of Physics and Research Laboratory of Electronics, Massachusetts Institute of Technology, Cambridge, MA 02139, USA
[4]Department of Material Science, University of Milano-Bicocca, Via Cozzi 55, 20121, Milano, Italy

kaminer@technion.ac.il

The complex range of interactions between electrons and electromagnetic fields gave rise to countless scientific and technological advances. A prime example is photon-induced nearfield electron microscopy (PINEM), enabling the detection of confined electric fields in illuminated nanostructures with unprecedented spatial resolution. However, PINEM is limited by its dependence on strong fields, making it unsuitable for sensitive samples, and its inability to resolve complex phasor information. Here, we leverage the nonlinear, over-constrained nature of PINEM to present an algorithmic microscopy approach, achieving far superior nearfield imaging capabilities. Our algorithm relies on free-electron Ramsey-type interferometry to produce orders-of-magnitude improvement in sensitivity and ambiguity-immune nearfield phase reconstruction, both of which are optimal when the electron exhibits a fully quantum behavior. Our results demonstrate the potential of combining algorithmic approaches with novel modalities in electron microscopy, and may lead to various applications from imaging sensitive biological samples to performing full-field tomography of confined light.

# I. Introduction

The interaction of electrons with static and dynamic electromagnetic fields lies at the center of numerous discoveries and applications. This interaction serves as the basic principle behind achieving atomic resolution [1,2]; it can result in new methods for characterizing the electronic [3] and magnetic [4] properties of materials; allows to promote and explore chemical processes [5]; and help understanding many key cellular functions by visualizing macromolecular machines [6]. In all above-mentioned examples, the nontrivial and multifaceted nature of the interaction is essential.

A prime example for the interaction of electrons with electromagnetic fields is photon-induced nearfield electron microscopy (PINEM) [7-9]. PINEM is an imaging technique relying on the inelastic scattering of free-electrons from illuminated structures to reconstruct nearfield amplitudes on the nanoscale with potentially sub-nm [10] and sub-ps [11] spatiotemporal resolutions. Beyond the time-resolved imaging of field dynamics [12], PINEM enables a plethora of additional abilities [13-18] such as detecting quantum emitter decoherence [19], reconstructing the quantum state of free electrons [20], generating attosecond electron bunches [21], and performing free electron wavefront shaping [22, 23].

However, when used for imaging, PINEM extracts only the amplitude of the nearfield without any information about its phase, and routinely neglects the inherent nonlinear connection between the nearfield and electron distributions [13, 14, 23]. Therefore, it is clear that a large amount of latent information still exists in PINEM and can be exploited for the efficient electromagnetic investigation of nanoscale objects including weakly-interacting ones. This can only be achieved by integrating PINEM with another methodology, to increase the possible degrees of freedom for investigation.

Notably, pre-modulating free electrons with a reference field has already proven to make them sensitive to the electron's wave phase [20, 21, 24, 25], dubbed as Ramsey-type phase control of free electrons [24]. Though sharing similar names, this phenomenon fundamentally differs from conventional, atom-based Ramsey interferometry schemes [26, 27], primarily since free electrons do not have a fixed set of internal degrees of freedom. Hence, the evolution of their quantum state under an applied AC electric field involves only a change in their total kinetic energy via photon absorption/emission [8], therefore applicable at arbitrary frequencies [28] and with a diverse range of samples.

More importantly, it boasts a large, multi-dimensional parameter space, including the intensities and relative phase of the applied reference and sample fields, as well as the spectral and temporal properties of the fields and the free electron itself. Thus, integrating Ramsey-type control with PINEM is highly suitable for an algorithmic approach to optimize the extracted sample information, while potentially gaining added advantage from the quantum dynamics the electron undergoes by changing its kinetic energy.

Here, we propose Free Electron Ramsey-type Imaging (FERI) to achieve enhanced imaging of both amplitude and phase of electromagnetic nearfields. We develop an algorithmic approach that optimizes the signal extracted from the over-constrained interaction, showing imaging with orders-of-magnitude less illumination power compared to conventional PINEM. We exemplify the approach by simulating FERI with illuminated gold nanospheres, which are potential markers in bioimaging [29, 30] (Fig. 1). FERI also enables ambiguity-immune phase-resolved nearfield imaging, requiring no prior knowledge regarding the sample, which we demonstrate by simulating the interaction of electrons with a hexagonal array of plasmonic vortices [23, 31].

Instead of relying on *transverse* electron coherence, as in electron holography methods [32, 33], FERI relies on *temporal* (longitudinal) coherence, greatly simplifying the electron source and optics, at the expense of more complex electromagnetic optics and addressing scheme. While our scheme is compatible with both quantum and classical, point-particle electrons, we provide analytical and numerical evidence that it performs optimally when the free electrons behave entirely as quantum particles. Thus, our approach promotes the use of light-driven electron microscopy in a plethora of platforms that could not have been investigated before, and allows sensitive, low-dose characterization of materials, gaining advantage from the quantum nature of the free electron.

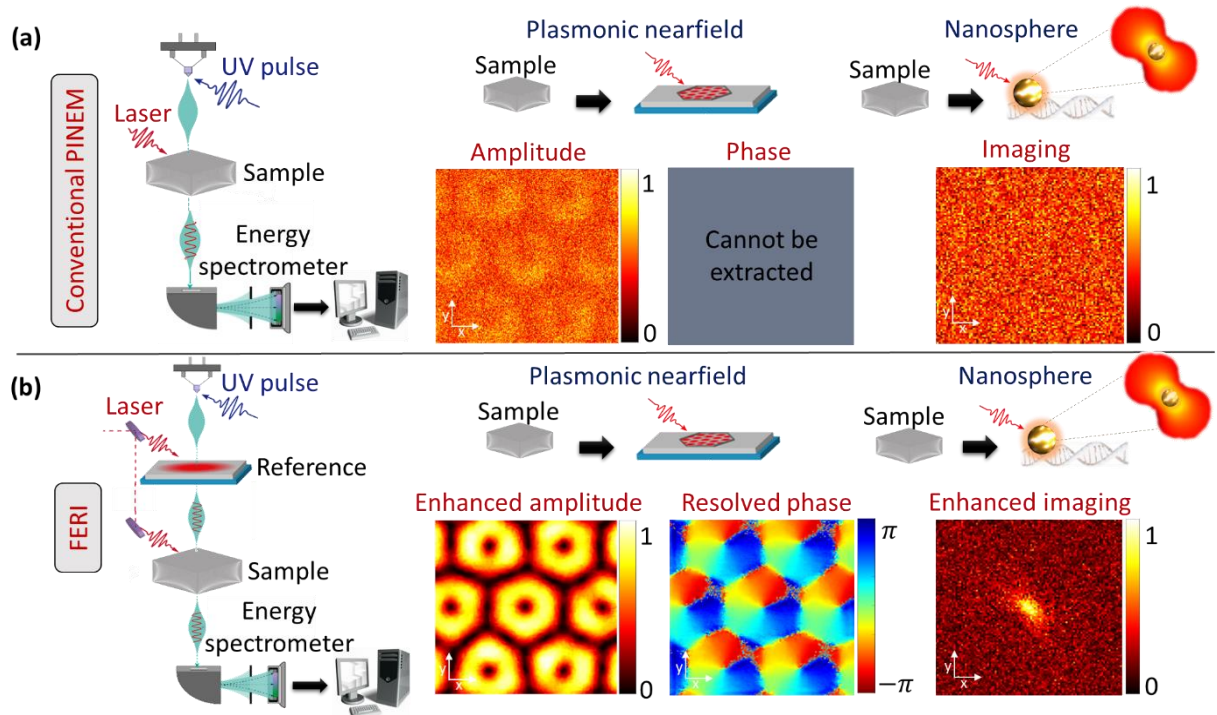

**Fig. 1 Free-electron Ramsey-type imaging (FERI): leveraging the complex free-electron interaction with confined fields for enhanced imaging.** (**a**) Conventional PINEM imaging versus (**b**) FERI. Left: The interaction configuration. Middle: Reconstruction of the amplitude and phase of plasmonic nearfields with hexagonal symmetry, at a similar signal-to-noise ratio and laser intensity. Right: Reconstruction of the weak nearfield induced by a gold nanosphere, which can be used as a marker for bioimaging.

## II. Imaging via pre-modulated electron pulses (FERI)

We begin by describing the interaction of electrons with nearfields – the theory that lies at the core of PINEM. The conventional PINEM theory usually considers an approximately monoenergetic electron of mean energy $E_0$ that interacts with a nearfield of central frequency $\omega$ [7, 8, 9, 11, 24]. Additionally, it is assumed that the electron energy spread $\sigma_E$, which is much smaller than $E_0$, is also smaller than the light quanta $\hbar\omega$, such that energy measurements distinguish between different number of absorbed/emitted quanta by the electron. This situation, commonly occurring in optical and near IR frequencies, is akin to requiring a quantum behavior of the electron. The majority of our analysis will focus on this case, but special attention will be given to the scenario $\sigma_E \gg \hbar\omega$ when analyzing the difference between a classical and quantum electron in FERI.

PINEM theory further assumes a paraxial electron with energy significantly higher compared to the light it interacts with ($E_0 \gg \hbar\omega$), such that the dispersion of the electron can be linearized. All the above-mentioned approximations are typically satisfied under the operating conditions of ultrafast scanning and transmission electron microscopes as shown in previous works in the field [7-9, 11, 24, 34-38].

In this case, the Hamiltonian describing the interaction between the electromagnetic field and the electron is given by:

$$H = E_0 - i\hbar v \partial_z + evE_z(x,y,z)/\omega. \tag{1}$$

Here $v$ is the velocity of the electron, $e$ is the fundamental charge, and $z$ is the direction of propagation of the electron. According to this Hamiltonian, interactions between the electron matter-wave and the electromagnetic wave result in a discrete energy exchange, such that the probability $P_l$ to measure the electron with energy $E_0 + l\hbar\omega$ (with $l$ being an integer) is given by $|J_l(2|g|)|^2$, where $g$ is a dimensionless interaction constant, $g = \frac{e}{\hbar\omega}\int_{-\infty}^{\infty} dz E_z(x,y,z)e^{-\frac{iz\omega}{v}}$. In general, $g = g(x,y)$ is a complex number and a function of the transverse coordinates $(x,y)$. By measuring $g(x,y)$, one can in principle recover both the amplitude and the phase of the nearfield $E_z(x,y)$, which in certain cases can suffice to reconstruct the full field information [31]. However, any energy filtered spectrum measurement $\sum_{l \in L_{\text{filter}}} P_l$ can only extract information about the absolute value of the interaction constant, $|g|$.

This is the case in conventional measurement schemes (Fig. 1a), where one filters the signal over the entire gain side ($l > 0$) to extract the nearfield at the sample, given by the interaction constant $g_s(x,y)$. In this case, we can use identities of Bessel functions [39] to conclude (see section SI2 in the Supp. Information for full derivation):

$$\text{Signal}(x,y) \propto \sum_{l=1}^{\infty} P_l(x,y) = \frac{1}{2}(1 - J_0(2|g_s(x,y)|)^2) \approx |g_s(x,y)|^2 \qquad (2)$$

In the weak field regime ($g \ll 1$), the signal is directly proportional to the field intensity. The quadratic dependence requires relatively intense fields to collect enough signal, making conventional PINEM inappropriate for sensitive samples.

Since $g$ is linear in the electric field, it can contain the superposition of two spatially separated fields, one acting as a reference and the other is the unknown sample field [22, 24, 25]. Fig. 1b presents the FERI scheme: it relies on the electron temporal (longitudinal) coherence to improve the detected PINEM signal at the sample by exploiting the additional coherent interaction with the reference field (with interaction constant $g_r$).

Fig. 2 shows the working mechanics of the FERI scheme for phase reconstruction. By adding an additional sub-cycle phase delay $\Delta\phi$ between the reference and sample fields, it is possible to extract both the amplitude and phase of the sample field using energy filtering on the post-interaction electron. The relative phase delay over-constrains the measurement, as it adds up several images taken with different relative phases. The different raw measurements are sampled from different distributions, thus enhancing image contrast rather than just improving SNR (as would happen in conventional PINEM if more images were to be added up).

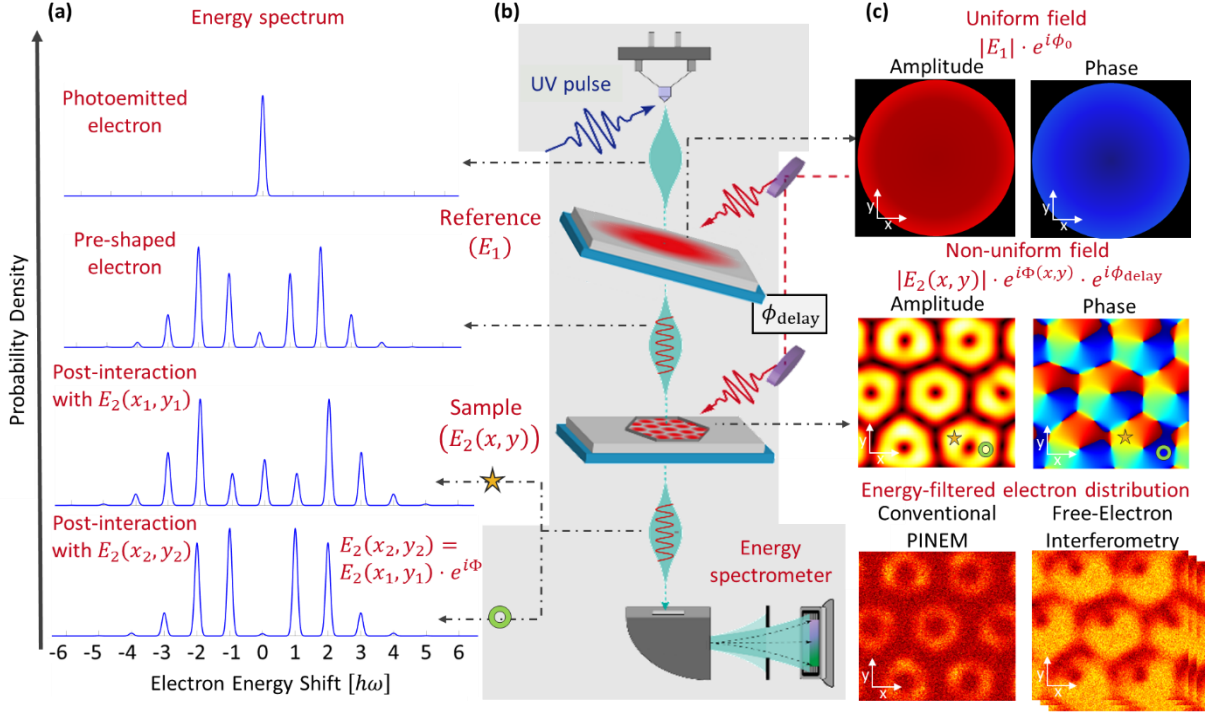

**Fig. 2 Free-electron Ramsey-type imaging measurement scheme.** A free electron in an ultrafast transmission electron microscope is pre-modulated via an interaction with a reference field. The pre-modulated electron probes the sample field, which is delayed by less than an optical cycle relative to the reference field. The complex value of the field (both amplitude and phase) is probed by measuring the distribution of energy-filtered electrons. **(a)** Evolution of the electron spectral probability density (SPD) as it propagates through the system, in given coordinates $(x, y)$. The bottom two SPD showcase how the phase information of the field and the nonlinear connection between its value and the electron distribution changes the measured signal at different locations. **(b)** Illustration of the FERI scheme. **(c)** Top: the reference field, consisting of a uniform amplitude and phase. Middle: the sample field (amplitude and phase). Bottom: energy-filtered electron distribution measurements for conventional PINEM (left) and FERI (right). From the latter, both the amplitude and phase can be reconstructed.

In the particularly interesting case of samples characterized by weak fields, we can replace Eq. 2 using a modified formula that takes directly into account the presence of a reference field (see section SI2 in the Supporting Information for full derivation):

$$\text{Signal}(x,y) \propto \sum_l P_l \approx \frac{1}{2}(1 - J_0(2|g_r|)^2) + 2J_0(2|g_r|)J_1(2|g_r|)|g_s(x,y)|\cos(\angle g_s(x,y) - \Delta\phi) \quad (3)$$

A critical point in Eq. 3 is the fact that the dynamic range of the measurement (in the limit of $|g_s| \ll 1$) scales as $|g_s|$, rather than of $|g_s^2|$. This fact alone results in significant enhancement of the sensitivity of FERI when compared to conventional PINEM. Furthermore, using Eq.3, we find the optimal amplitude of the reference field interaction constant ($|g_r^{\text{optimal}}|$) *for any*

*given case where* $|g_s| \ll 1$. It is given by the first positive solution to the equation $J_0(2x)^2 - J_2(2x)J_0(2x) - 2J_1(2x)^2 = 0$ (see SI2), which yields $|g_r^{\text{optimal}}| \approx 0.541$ – a value well within reach of typical PINEM experiments [11]. Using this optimal reference interaction, the dynamic range of the signal in a sample with maximal interaction $|g_s^m| \ll 1$ increases by the factor of $2J_1(2|g_r^{\text{optimal}}|)J_0(2|g_r^{\text{optimal}}|)/|g_s^m| \sim 0.678/|g_s^m|$ (see SI2). It is noteworthy that more degrees of freedom arise from this analysis, such as scanning over different reference field interaction constants $|g_r|$ or by introducing and varying free-space propagation (see SI1).

### III. Enhanced and phased-resolved nearfield imaging

PINEM applications for field imaging have thus far used energy-filtered transmission electron microscopy (EFTEM), filtering electrons over a certain range, only extracting part of the spatial information of the field. The signal in such cases is linear in the field intensity. Once we perform a pre-modulating interaction, the field at the sample can interfere with different interaction orders $l$, and by energy filtering the electrons, the nonlinearity of the relation between field and electron distributions provides a substantial additional information.

The general energy-filtered electron distribution measurement can be described as $M = \sum_{l \in L_{\text{filter}}} P_l = \sum_{l \in L_{\text{filter}}} |J_l(2|g_{\text{total}}|)|^2$, with $g_{\text{total}} = g_r + g_s$ and $L_{\text{filter}}$ being the range of filtered energies. The measured signal for each relative sub-cycle delay $\Delta\phi$ and each transverse sample coordinate $x, y$, is:

$$M(x, y, \Delta\phi, L_{\text{filter}}, |g_s(x,y)|, \angle g_s(x,y)) =$$

$$\sum_{l \in L_{\text{filter}}} \left| J_l\left(2\sqrt{|g_r|^2 + |g_s(x,y)|^2 + 2|g_r||g_s(x,y)|\cos(\angle g_s(x,y) - \Delta\phi)}\right) \right|^2.$$

The measurement model expression is ambiguous, i.e., multiple values of $\{|g_s|, \angle g_s\}$ can output the same value of energy-filtered electron distribution measurement, denoted as $Y(x, y, \Delta\phi, L_{\text{filter}})$. To remove this ambiguity and extract the correct field $g_s$, the optimization procedure (Fig. 3) scans over the relative phase between the reference field and sample field. Alhough not used in this work, additional information can also be gained by scanning over the filtered energy range. By using maximum likelihood estimation (MLE) with the measurement model $M$, the reconstruction of the amplitude and phase of the sample is performed per $\{x, y\}$ coordinate by minimizing the following expression:

$$\underset{|g_s(x,y)|, \angle g_s(x,y)}{\mathrm{argmin}} \sum_{\Delta\phi, L_{\mathrm{filter}}} |Y(x, y, \Delta\phi, L_{\mathrm{filter}}) - M(x, y, \Delta\phi, L_{\mathrm{filter}}, |g_s(x,y)|, \angle g_s(x,y))|^2. \quad (4)$$

The measurement scheme can also be more generally performed with scanning TEM electron energy loss spectroscopy (STEM-EELS), where the entire electron energy spectrum is measured per point, as opposed to EFTEM, which measures the entire spatial distribution for a given electron energy range. Performing STEM-EELS (which acts similarly to performing EFTEM while scanning over the energy filtered range) may improve the field reconstruction, but requires longer data acquisition times.

As the MLE expression in Eq. 4 is not convex, gradient descent can converge to a local minimum. To solve this issue, the minimization method used is summation of MLE heatmaps for all relative phases (and energy filters, if relevant), while choosing the minimum value of the joint MLE heatmap. This approach guarantees convergence to the global minima, as exemplified in Fig. 3 for an arbitrary phase profile and for plasmonic optical vortices [31]. After reconstructing $|g_s(x, y)|, \angle g_s(x, y)$ for every coordinate $\{x, y\}$, we further use a denoising convolutional neural network (DnCNN) [40] for further signal improvement, making use of natural image prior knowledge. The DnCNN utilizes the concept of residual learning, making it possible to denoise arbitrary confined field images.

Many phase retrieval methods require solving optimization problems for the transverse plane, but they often encounter ambiguities such as global phase shifts, conjugate inversions, and spatial shifts. More complex ambiguities stem from the non-convexity of the optimization problem [41, 42]. In our example of plasmonic optical vortices, determining the rotation direction of the vortex presents an ambiguity. However, FERI prevents these ambiguities since its optimization is performed per pixel instead of on the entire transverse plane, while utilizing combinations of several MLE heatmaps.

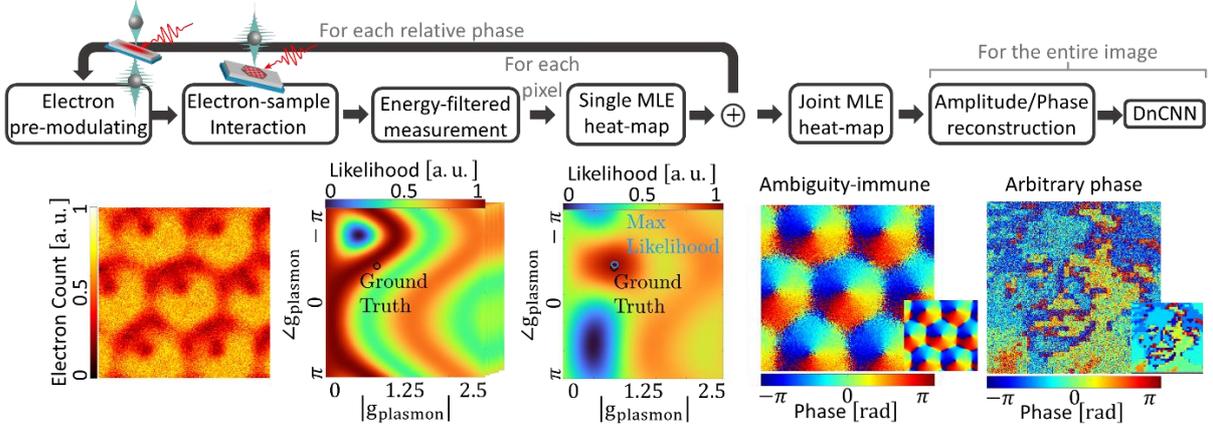

**Fig. 3 Phase-resolved optimization.** Block scheme (top) and visualization (bottom) of the optimization process. The free electrons are pre-modulated by a reference field before they probe the sample field and are energy-filtered to produce electron distribution measurements (bottom row, 1st from the left). For each transverse pixel, several MLE heat-maps are generated, for different relative phases between the reference and signal fields (bottom row, 2nd from the left). The summed joint MLE heat-map (bottom row, 3rd from the left) enables the accurate estimation of the amplitude and phase for each pixel. Finally, a denoising convolutional neural network (DnCNN) enhances the amplitude/phase reconstruction. To the right are two phase-reconstruction examples: plasmonic optical vortices and an arbitrary phase (a pixelated Einstein portrait), with the corresponding ground truth images in insets.

We exemplify this procedure on plasmonic standing waves with hexagonal geometry (Fig.2c), which can be measured in PINEM experiments [23]. We compare our FERI scheme with conventional PINEM using the structural similarity index measure (SSIM) [43] and show an improvement by almost two orders of magnitude for the minimal interaction strength necessary for the reconstruction (Fig.4a).

We derive the minimal interaction strength needed to successfully image a given SNR. For conventional PINEM, it follows $\left|g_{min}^{PINEM}\right| \approx 10^{-\frac{SNR_{dB}}{40}-1.2}$; when with FERI – $\left|g_{min}^{FERI}\right| \approx 10^{-\frac{SNR_{dB}}{19.7}-2.2}$, and when introducing DnCNN – $\left|g_{min}^{DnCNN-FERI}\right| \approx 10^{-\frac{SNR_{dB}}{19.7}-2.5}$. The parameters were extracted from a linear fit to the data in Fig.4a and can vary by the choice of

SSIM threshold. That said, the difference in scaling follows the analytical relation for the minimal interaction strength:

$$|g_{\min}^{\text{PINEM}}| \approx \frac{1}{\sqrt{2}} \cdot \sqrt{|g_{\min}^{\text{FERI}}|} \approx \sqrt{2} \cdot \sqrt{|g_{\min}^{\text{DnCNN-FERI}}|} \qquad (5)$$

This square relation between PINEM and FERI is consistent with Eq.2 and Eq.3, while the factor of $\frac{1}{\sqrt{2}}$ relates to the dynamics range derived in section II. We can see here that by using further algorithmic improvements, such as DnCNN, we can improve the multiplying factor to the minimal interaction strength.

Such improvements in sensitivity and low-dose operation are especially important when attempting to image sensitive materials or objects weakly interacting with the applied electromagnetic field. To quantify these prospects, we analyze the FERI-based enhancement of PINEM signals from gold nanoparticles, which can be used as markers in bioimaging, sequencing, and diagnosis [29, 44-47]. By using similar laser intensity and SNR, as the current state of the art in low-dose PINEM [14], we demonstrate that conventional PINEM can detect gold nanospheres with a radius of 30 nm, while detection down to 14 nm is possible with FERI (Fig.4b). From the perspective of power efficiency, we show 100-fold reduction in necessary laser fluence, down to $\sim 0.1 \left[\frac{\mu J}{cm^2}\right]$.

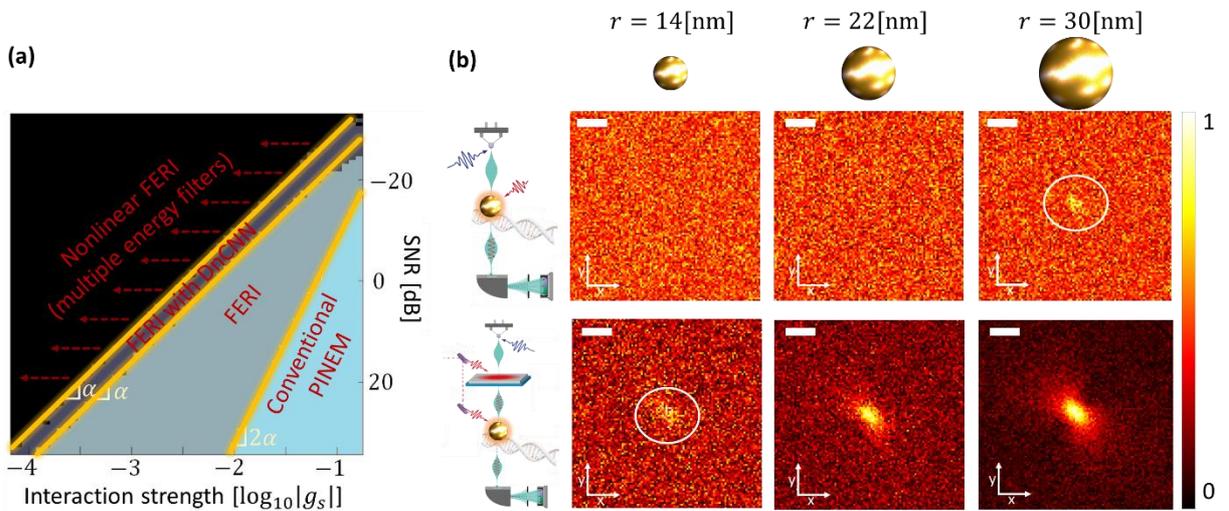

**Fig. 4. Enhanced nearfield imaging using FERI.** (a) Successful amplitude reconstruction (SSIM > 0.4) for conventional PINEM (turquoise), FERI (light gray), and with an added

DnCNN (dark gray). Additional information, such as that gained by using several energy filtering ranges, increases sensitivity even further. **(b)** Laser-excited gold nanospheres imaging simulations for different sphere radii. Top: Conventional PINEM imaging, Bottom: FERI for the same noise values, number of measurements, and excitation intensity. Scale bar is 15 nm. The spheres are envisioned as markers for biological samples, e.g. single DNA strings.

## IV. The quantum nature of free electrons and its effect on FERI

Up until now, we only considered the electron as a quantum particle, i.e. that its energy spread $\sigma_E$ is smaller than the light energy quantum $\hbar\omega$. Indeed, in the limit $g_s \ll g_r \ll 1$, Eq. 3 seems to directly reproduce the result of using the quantum coherence of a free-electron for sensing applications [48], though some changes to the measured signal are required. Nevertheless, the electron-mediated interference term between the fields in Eq. 3 results solely from *their* coherence, as well as the persisting temporal coherence of the electron. Hence, even a classical electron should gain significant improvement in sensitivity by using our scheme, with information stored in the classical analogue of the quantum phase – the electron velocity.

A classical, point particle electron can be directly defined by the limit $\sigma_E \gg \hbar\omega$, and it produces a conventional PINEM signal only when $g\hbar\omega > \sigma_E$, requiring a substantial investment of energy ($g \gg 1$). This is a direct result of the inability to completely separate the initial and light-induced electron spectral probability density, or more generally – to resolve specific PINEM orders. Though FERI should improve sensitivity, the interference term within the PINEM signal, which is responsible for this improvement, is reduced by the factor $\mathrm{erf}\left(\frac{\hbar\omega}{\sigma_E}\right)$ [48].

The direct consequence of this reduced sensitivity is a rescaling of interaction constants, increasing both the optimal reference interaction constant and the required sample interaction constant by the same factor, $\mathrm{erf}\left(\frac{\hbar\omega}{\sigma_E}\right)$. Thus, though maximal FERI-induced enhancement to PINEM remains the same, the minimal possible energy is invested only when the electron acts as a quantum particle.

**V. Discussion and outlook**

The enhancement of FERI fundamentally arises from the longitudinal (temporal) coherence of the electron (either quantum or classical). It is intriguing to compare this approach with more conventional methods that rely on transverse coherence, such as electron holography [32], which were recently studied with ultrafast electron microscopes [22, 33, 49]. Utilizing longitudinal coherence avoids limitations due to electron beam quality and could enable operation with simpler, higher-flux electron sources. Since longitudinal electron coherence represents a robust parameter, FERI measurements can even be performed with slower electrons in scanning electron microscopes (SEMs) [50]. Thus, FERI seems to be a route that is easier to follow.

possible extensions of our work can explore increasing the spatial resolution to the atomic scale, employing sample tilt to perform full-field tomography, and operate at low (<10K) temperatures using liquid helium cryo-holders [51], providing a unique probe for cold condensed matter physics. The sensitivity enhancement showed in this work also makes FERI a promising approach to demonstrate enhanced cathodoluminescence [52] as well as free-electron-bound-electron resonant interaction (FEBERI) [53], which had not been realized experimentally so far, due to its intrinsically weak interaction.

**References**

[1] K. W. Urban, *Studying Atomic Structures by Aberration-Corrected Transmission Electron Microscopy,* Science 321, 5888, 506-510 (2008).

[2] A. Polman, M. Kociak, and F. J. García de Abajo, *Electron-Beam Spectroscopy for Nanophotonics*, Nat. Mater. 18, 1158 (2019).

[3] H. J. Leamy, *Charge collection scanning electron microscopy*. J. Appl. Phys. 53, R51–R80 (1982).

[4] X. Fu, S. D. Pollard, B. Chen, B. Yoo, H. Yang and Y. Zhu, *Optical manipulation of magnetic vortices visualized in situ by Lorentz electron microscopy,* Science Advances 4, 7 (2018).

[5] J. M. Thomas and O. Terasaki, *The Electron Microscope Is an Indispensable Instrument for the Characterisation of Catalysts*, Topics in Catalysis 21, 155–159 (2002).

[6] W. Baumeister and A. C. Steven, *Macromolecular electron microscopy in the era of structural genomics*, Trends in biochemical sciences 25, 12, 624-631 (2000).

[7] B. Barwick, D. J. Flannigan, and A. H. Zewail, *Photon-Induced Near-Field Electron Microscopy*, Nature 462, 902 (2009).

[8] F. J. García De Abajo, A. Asenjo-Garcia, and M. Kociak, *Multiphoton Absorption and Emission by Interaction of Swift Electrons with Evanescent Light Fields*, Nano Lett. 10, 1859 (2010).

[9] S. T. Park, M. Lin, and A. H. Zewail, *Photon-Induced Near-Field Electron Microscopy (PINEM): Theoretical and Experimental*, New J. Phys. 12, 123028 (2010).

[10] A. H. Zewail, *Four-Dimensional Electron Microscopy*, Science 328, 187 (2010).

[11] A. Feist, K. E. Echternkamp, J. Schauss, S. V. Yalunin, S. Schäfer, and C. Ropers, *Quantum Coherent Optical Phase Modulation in an Ultrafast Transmission Electron Microscope*, Nature 521, 200 (2015).

[12] Y. Kurman, R. Dahan, H. H. Sheinfux, K. Wang, M. Yannai, Y. Adiv, O. Reinhardt, L. H. G. Tizei, S. Y. Woo, J. Li, J. H. Edgar, M. Kociak, F. H. L. Koppens, I. Kaminer, *Spatiotemporal imaging of 2D polariton wave packet dynamics using free electrons*. Science 372, 1181–1186 (2021).

[13] O. Kfir, H. Lourenço-Martins, G. Storeck, M. Sivis, T. R. Harvey, T. J. Kippenberg, A. Feist, and C. Ropers, *Controlling Free Electrons with Optical Whispering-Gallery Modes*, Nature 582, 46 (2020).

[14] K. Wang, R. Dahan, M. Shentcis, Y. Kauffmann, A. Ben Hayun, O. Reinhardt, S. Tsesses, and I. Kaminer, *Coherent Interaction between Free Electrons and a Photonic Cavity*, Nature 582, 50 (2020).

[15] G. M. Vanacore, G. Berruto, I. Madan, E. Pomarico, P. Biagioni, R. J. Lamb, D. McGrouther, O. Reinhardt, I. Kaminer, B. Barwick, H. Larocque, V. Grillo, E. Karimi, F. J.

García de Abajo, and F. Carbone, *Ultrafast Generation and Control of an Electron Vortex Beam via Chiral Plasmonic Near Fields*, Nat. Mater. 18, 573 (2019).

[16] T. T. A. Lummen, R. J. Lamb, G. Berruto, T. Lagrange, L. Dal Negro, F. J. García De Abajo, D. McGrouther, B. Barwick, and F. Carbone, *Imaging and Controlling Plasmonic Interference Fields at Buried Interfaces*, Nat. Commun. 7, 13156 (2016).

[17] L. Piazza, T. T. A. Lummen, E. Quiñonez, Y. Murooka, B. W. Reed, B. Barwick, and F. Carbone, *Simultaneous Observation of the Quantization and the Interference Pattern of a Plasmonic Near-Field*, Nat. Commun. 6, 6407 (2015).

[18] Y. Adiv, K. Wang, R. Dahan, P. Broaddus, Y. Miao, D. Black, K. Leedle, R. L. Byer, O. Solgaard, R. J. England, and I. Kaminer, *Quantum Nature of Dielectric Laser Accelerators*, Phys. Rev. X 11, 041042 (2021).

[19] R. Ruimy, A. Gorlach, C. Mechel, N. Rivera, and I. Kaminer, *Towards atomic-resolution quantum measurements with coherently-shaped free electrons*. Phys. Rev. Lett. 126, 233403 (2021).

[20] KE. Priebe, C. Rathje, S. V. Yalunin, T. Hohage, A. Feist, S. Schäfer and C. Ropers, *Attosecond electron pulse trains and quantum state reconstruction in ultrafast transmission electron microscopy*. Nat. Phot. 11, 793 (2017).

[21] Y. Morimoto, and P. Baum, *Diffraction and microscopy with attosecond electron pulse trains*. Nat. Phys. 14, 252–256 (2018).

[22] I. Madan, G. M. Vanacore, E. Pomarico, G. Berruto, R. J. Lamb, D. McGrouther, T. T. A. Lummen, T. Latychevskaia, F. J. García de Abajo, F. Carbone, *Holographic imaging of electromagnetic fields via electron-light quantum interference.* Sci. Adv. 5, eaav8358 (2019).

[23] S. Tsesses, R. Dahan, K. Wang, T. Bucher, K. Cohen, O. Reinhardt, G. Bartal, and I. Kaminer, *Tunable Photon-Induced Spatial Modulation of Free Electrons*, Nat. Mater. (2023).

[24] K. E. Echternkamp, A. Feist, S. Schäfer, C. Ropers, *Ramsey-type phase control of free-electron beams*. Nat. Phys. 12, 1000–1004 (2016).

[25] A. Ryabov, J. W. Thurner, D. Nabben, M. V. Tsarev, and P. Baum, *Attosecond metrology in a continuous-beam transmission electron microscope*. Sci. Adv. 6, eabb1393 (2022).

[26] N. Ramsey, *Molecular Beams*, Vol. 20 (Oxford University Press, 1956).

[27] P. Bertet, S. Osnaghi, A. Rauschenbeutel, G. Nogues, A. Auffeves, M. Brune, J. M. Raimond, and S. Haroche, *A Complementarity Experiment with an Interferometer at the Quantum-Classical Boundary*, Nature 411, 166 (2001).

[28] O. Kfir, V. Di Giulio, F. J. G. de Abajo, C. Ropers, *Optical coherence transfer mediated by free electrons*. Sci. Adv. 7, eabf6380 (2021).

[29] R. A. Sperling, P. R. Gil, F. Zhang, M. Zanella and W. J. Parak, *Biological applications of gold nanoparticles*. Chem. Soc. Rev. 37, 1896–1908 (2008).

[30] M. De, P. S. Ghosh, V. M. Rotello, *Applications of Nanoparticles in Biology,* Advanced Materials 20, 4225–4241 (2008).

[31] S. Tsesses, K. Cohen, E. Ostrovsky, B. Gjonaj, G. Bartal, *Spin-Orbit Interaction of Light in Plasmonic Lattices*. Nano Lett. 19, 4010–4016 (2019).

[32] P. Simon, H. Lichte, P. Formanek, M. Lehmann, R. Huhle, W. Carrillo-Cabrera, A. Harscher, H. Ehrlich, *Electron holography of biological samples*. Micron 39, 229–256 (2008).

[33] J. H. Gaida, H. Lourenço-Martins, S. V. Yalunin, A. Feist, M. Sivis, T. Hohage, J. García de Abajo, C. Ropers. *Lorentz Microscopy of Optical Fields*. Research Square (2022).

[34] O. Reinhardt and I. Kaminer, *Theory of Shaping Electron Wavepackets with Light,* ACS Photonics 7, 10, 2859–2870 (2020).

[35] S. T. Park, A. H. Zewial, *Relativistic Effects in Photon-Induced Near Field Electron Microscopy,* Journal of Physical Chemistry A, 116 (46), 11128-11133 (2012).

[36] S. T. Park, A. H. Zewial, *Photon-induced near-field electron microscopy: mathematical formulation of the relation between the experimental observables and the optically driven charge density of nanoparticles,* Physical Review A: Atomic, Molecular, and Optical Physics, 89, 013851-013851 (2014).

[37] R. Shiloh, Y. Lereah, Y. Lilach and A. Arie, *Sculpturing the electron wave function using nanoscale phase masks*. Ultramicroscopy 144, 26–31 (2014).

[38] D. Roitman, R. Shiloh, P. H. Lu, R. E. Dunin-Borkowski and A. Arie, *Shaping of electron beams using sculpted thin films*. ACS Photon. 8, 3394–3405 (2021).

[39] F. W. Olver, D. W. Lozier, R. F. Boisvert and C. W. Clark, *NIST handbook of mathematical functions hardback and CD-ROM* (Cambridge university press, 2010).

[40] K. Zhang, W. Zuo, Y. Chen, D. Meng and L. Zhang, *Beyond a Gaussian denoiser: residual learning of deep CNN for image denoising*. IEEE Trans. Image Process 26, 3142–3155 (2017).

[41] Y. Shechtman, Y. C. Eldar, O. Cohen, H. N. Chapman, J. Miao and M. Segev, *Phase Retrieval with Application to Optical Imaging: A contemporary overview*, in IEEE Signal Processing Magazine,32 3, 87-109, (2015).

[42] L. Taylor, *The phase retrieval problem*, in IEEE Transactions on Antennas and Propagation, 29, 2, 386-391, (1981).

[43] Z. Wang, A. C. Bovik, H. R. Sheikh and E. P. Simoncelli, *Image quality assessment: from error visibility to structural similarity*. IEEE Trans. Image Process. 13, 600–612 (2004).

[44] Y. Wu, M.R. Ali, K. Chen, N. Fang, M.A. El-Sayed, *Gold nanoparticles in biological optical imaging*, Nano Today, 24, 120-140 (2019).

[45] L. Novotny and N. van Hulst, *Antennas for light*. Nature Photon. 5, 83–90 (2011)

[46] J. J. Storhoff, A. D. Lucas, V. Garimella, Y. P. Bao and U. R. Müller, *Homogeneous detection of unamplified genomic DNA sequences based on colorimetric scatter of gold nanoparticle probes*. Nature Biotechnol. 22, 883–887 (2004).

[47] S. Sargazi, U, Laraib, S. Er, *Application of green gold nanoparticles in cancer therapy and diagnosis*. Nanomaterials 12(7), 1102 (2022).

[48] A. Karnieli, S. Tsesses, R. Yu, N. Rivera, Z. Zhao, A. Arie, S. Fan and I. Kaminer, *Quantum sensing of strongly coupled light-matter systems using free electrons,* Sci. Adv. 9, eadd2349 (2023).

[49] F. Houdellier, G.M. Caruso, S. Weber, M.J. Hÿtch, C. Gatel, A. Arbouet, *Optimization of off-axis electron holography performed with femtosecond electron pulses,* Ultramicroscopy, 202, 26-32 (2019).

[50] R. Shiloh, T. Chlouba and P. Hommelhoff, *Quantum-Coherent Light-Electron Interaction in a Scanning Electron Microscope*, Phys. Rev. Lett. 128, 235301 (2022).

[51] K. A. Taylor, and R. M. Glaeser, *Electron diffraction of frozen, hydrated protein crystals*. Science 186, 1036–1037 (1974).

[52] M. Taleb, M. Hentschel, K. Rossnagel, H. Giessen and N. Talebi, *Phase-locked photon–electron interaction without a laser*, Nat. Phy. (2023).

[53] A. Gover and A. Yariv, *Free-Electron-Bound-Electron Resonant Interaction*, Phys. Rev. Lett. 124, 64801 (2020).